\documentclass[aps,prl,twocolumn,superscriptaddress]{revtex4-2}
\usepackage{graphicx}
\usepackage{amsmath}
\usepackage{amssymb}
\usepackage{colordvi}
\usepackage{mathrsfs}
\usepackage{bm}
\usepackage{verbatim}
\usepackage{dcolumn}
\usepackage{bm}
\usepackage{epsfig}
\usepackage{subfigure}
\usepackage{multirow}
\usepackage[colorlinks=true,allcolors=blue]{hyperref}
\usepackage{pifont}

\begin{document}
\title{The Layer Hall Effect without External Electric Field}

\author{Yulei Han}
\email[Correspondence author:~~]{han@fzu.edu.cn}
\affiliation{Department of Physics, Fuzhou University, Fuzhou, Fujian 350108, China}

\author{Yunpeng Guo}
\affiliation{Department of Physics, Fuzhou University, Fuzhou, Fujian 350108, China}

\author{Zeyu Li}
\affiliation{International Center for Quantum Design of Functional Materials,
CAS Key Laboratory of Strongly-Coupled Quantum Matter Physics, and Department of Physics,
University of Science and Technology of China, Hefei, Anhui 230026, China}
\affiliation{Hefei National Laboratory, University of Science and Technology of China, Hefei 230088, China}

\author{Zhenhua Qiao}
\email[Correspondence author:~]{qiao@ustc.edu.cn}
\affiliation{International Center for Quantum Design of Functional Materials,
CAS Key Laboratory of Strongly-Coupled Quantum Matter Physics, and Department of Physics,
University of Science and Technology of China, Hefei, Anhui 230026, China}
\affiliation{Hefei National Laboratory, University of Science and Technology of China, Hefei 230088, China}
\date{\today{}}

\begin{abstract}
   The layer Hall effect is an intriguing phenomenon observed in magnetic topological layered materials, where the Hall response arises from the opposite deflection of electrons on top and bottom layers.
   To realize layer Hall effect, space-time $\mathcal{PT}$ symmetry is typically broken by applying an external electric field. In this work, we propose a new mechanism to realize the layer Hall effect by introducing inequivalent exchange fields on both surfaces of a topological insulator thin film, in the absence of an electric field. This approach yields a distinct Hall response compared to the conventional electric-field-induced layer Hall effect, particularly with respect to the Fermi level. Taking the topological insulator Sb$_2$Te$_3$ as a concrete example, we demonstrate the feasibility of inducing the layer Hall effect only by coupling the top and bottom surfaces of Sb$_2$Te$_3$ with different magnetic insulators. Notably, we show that both built-in electric-field-induced and inequivalent exchange-fields-induced layer Hall effects can be achieved by tuning the stacking order between Sb$_2$Te$_3$ and the magnetic layers. Given the well-established experimental techniques for fabricating topological insulator thin films, our work offers a viable pathway for realizing layer Hall effect without external electric field.
 \end{abstract}

\maketitle

\textit{Introduction---.} The topological insulator (TI) is a fascinating state of matter that has been extensively explored over the past two decades~\cite{Hasan_2010,Qi_2011,Ren_2016,Breunig_2022}. Unlike conventional insulators, TIs are insulating in the bulk but host conducting helical surface states protected by time-reversal symmetry. When magnetism is introduced into TIs -- either through intrinsic magnetic order or doping with magnetic elements -- a novel class of topological materials, known as magnetic topological insulators, emerges with broken time-reversal symmetry~\cite{Tokura_2019,Bernevig_2022,Wang_2021,Liu_2023}. The interplay between magnetism and topology in magnetic TIs leads to exotic quantum phenomena, such as the quantum anomalous Hall effect~\cite{Liu_2016, Chang_2023,Yu_2010,Qiao_2010,Chang_2013,Qiao_2014,Wang_2014,Chang_2015,Qi_2016,Sun_2019,Petra_2020,Deng_2020,Zhao_2020,Serlin_2020,Chen_2020,Li_2021,Li_2022,Zhao_2022,Han_2023,Han_2024,Deng_2024} and axion insulators~\cite{Wang_2015,Morimoto_2015,Mogi_2017,Mogi_2017_2,Varnava_2018,Xiao_2018,Xu_2019,Liu_2020,Nenno_2020,Li_2023,Qin_2023,Li_2024}, which have been widely explored in various systems. Due to their nontrivial properties, magnetic TIs hold significant potential for applications in dissipationless electronic devices, quantum computing, and the topological magnetoelectric effect.

The intrinsic origin of these quantum phenomena lies in the distinct Berry curvature distributions within the first Brillouin zone. For instance, the quantum anomalous Hall effect typically observed in ferromagnetic systems exhibits nonzero Berry curvature, resulting in a nonzero Chern number that determines the number of chiral edge states. In contrast, an axion insulator with antiferromagnetic coupling features the top and bottom layers that contribute opposite Berry curvatures, leading to zero net Berry curvature due to the preservation of space-time $\mathcal{PT}$ symmetry. If the $\mathcal{PT}$ symmetry of an axion insulator is broken, such as by applying an external electric field, the layer-locked Berry curvature can induce a nonzero anomalous Hall conductivity. The phenomenon is known as the layer Hall effect (LHE) and was recently observed in the even-layered antiferromagnetic TI MnBi$_2$Te$_4$~\cite{Gao_2021}. Encoding Berry curvature with layer degree of freedom opens avenues for application in topological antiferromagnetic spintronics.

To date, the LHE has been proposed in various systems, including MnBi$_2$Te$_4$-family materials~\cite{Gao_2021,Chen_2022,Dai_2022,Peng_2023,Xu_2024}, ferroelectric~\cite{Zhang_2023,Zhang_2024} and multiferroic~\cite{Feng_2023,Liu_2024} van der Waals magnetic bilayers, and transition metal oxides~\cite{Tao_2024}.
In these systems, the realization of LHE typically requires the application of an external electric field~\cite{Gao_2021,Chen_2022,Dai_2022,Zhang_2024,Tao_2024} or the induction of built-in electric fields through mechanisms such as ferroelectric coupling~\cite{Xu_2024,Liu_2024} or layer sliding~\cite{Peng_2023, Zhang_2023,Feng_2023}.
However, aside from MnBi$_2$Te$_4$, the experimental implementation of these materials remains challenging due to the complexities of controlling van der Waals monolayer or bilayer structures~\cite{Cai_2018}, and the consistent need for an electric field. Therefore, it is crucial to seek a new physical mechanism and a practical approach compatible with current experimental techniques for realizing LHE.

In this Letter, we propose a new mechanism to realize the LHE by introducing inequivalent exchange fields on both surfaces of a TI thin film without applying an external electric field. By thoroughly exploring the differences between electric-field-induced and exchange-field-induced LHEs, we find that the distinct dependences of anomalous Hall conductivity on the Fermi level serves as a key hallmark for distinguishing between the two types of LHE. To further validate the feasibility of this approach, we construct a heterostructure consisting of a five quintuple-layer (QL) Sb$_2$Te$_3$ coupled with a monolayer magnetic insulator MnBi$_2$Te$_4$ (FeBi$_2$Te$_4$) on the bottom (top) surface. By using first-principles calculations, we demonstrate that the stacking order between Sb$_2$Te$_3$ and magnetic layers plays a crucial role in determining the type of LHE.

\begin{figure}[t]
  \centering
  \includegraphics[width=8.5 cm]{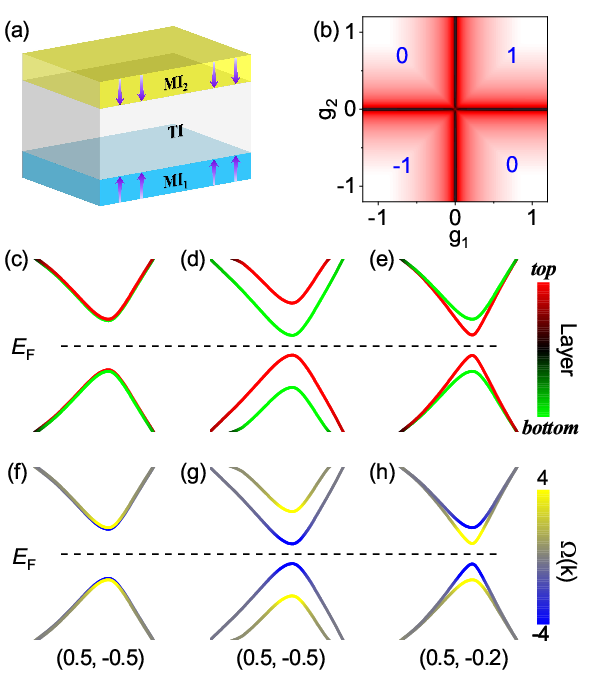}
  \caption{(a) Schematic plot of the sandwich heterostructure composed of magnetic insulator and topological insulator, with purple arrows denoting the spin direction. (b) Phase diagram in ($g_1$, $g_2$) spaces, with Chern numbers indicated in each insulating region. (c)-(h) Illustration of LHE in a 1-layer-MI$_1$/5-layer-TI/1-layer-MI$_2$ system. The first, second, and third columns correspond to the axion insulator, electric-field-induced LHE, and inequivalent exchange-field-induced LHE, respectively. (c)-(e) Layer-resolved and (f)-(h) Berry-curvature-resolved band structures for surface states. The exchange field strengths ($g_1$, $g_2$) are labeled in each column. }\label{Fig1}
\end{figure}

\textit{Lattice Models---.} To explore the LHE induced by inequivalent exchange fields, we construct a sandwich structure consisting of a TI thin film coupled with two magnetic insulators (MIs) on both surfaces, as shown in Fig.~\ref{Fig1}(a). The corresponding tight-binding model Hamiltonian can be expressed as~\cite{Fu_2007,Jiang_2012}
\begin{equation}\label{Ham}
  H = \sum_{\alpha,i}c_{i+1}^{\dagger} T_{\alpha}c_{i} + \sum_{i}(E_0 + m_{\rm{EX}})c_{i}^{\dagger}c_{i} + h.c.,
\end{equation}
where $T_{\alpha} = B\sigma_{z}\otimes s_{0}+\frac{iA}{2}\sigma_{x}\otimes s_{\alpha}$ with $\alpha = x,~y,~z$, $E_{0} = (M_{0}(l)-6B)\sigma_{z}\otimes s_{0}$, and the exchange field $m_{\rm{EX}} = m(l)\sigma_{0}\otimes s_{z}$ with a layer-dependent strength $m(l)$. $\mathbf{\sigma}$ and $\mathbf{s}$ are Pauli matrices for the orbital and spin degrees of freedom, respectively. The layer-dependent parameters $M_{0}(l)$ and $m(l)$ are defined as: $M_{0}(l)=M_{0}$ and $m(l)=0$ for TI layers; $M_{0}(l)=M_{i}$ and $m(l)=g_i$ for the $i$-th MI. For simplicity, we set $A=1.0$, $B=0.7$, $M_{0}=1.0$, and $M_{i}= 0.3$. It is noteworthy that the sign of $M_{i}$, which determines whether the MI is in a topological ($M_{i} >0$) or trivial ($M_{i} <0$) phase, does not qualitatively affect the topological phase diagrams in the MI-TI stacking systems~\cite{Deng_2024}. The layer-resolved Chern number $\mathcal{C}_{z}$ can be calculated by~\cite{Li_2024}:
\begin{equation}
  \mathcal{C}_{z} = \frac{1}{\pi}\sum_{E_{n}<E_{F}<E_{n'}} \int d\mathbf{k} \mathrm{Im} \frac{\langle u_{n\mathbf{k}}|\hat{P}_{z} v_x|u_{n'\mathbf{k}}\rangle \langle u_{n'\mathbf{k}}|v_y|u_{n\mathbf{k}}\rangle}{(E_{n'}-E_{n})^2}
\end{equation}
where $v_{x,y}$,  $u_n(\mathbf{k})$, and $\hat{P}_z$ represent the velocity operators, eigenstates, and the projecting operator, respectively. The summation runs over all occupied bands in the first Brillouin zone.

\begin{table}
\centering
\caption{Symmetry analysis of MI$_1$-TI-MI$_2$ heterostructure without/with the presence of electric field $E$.}\label{table1}
\setlength{\tabcolsep}{5mm}
\begin{tabular}{ccccc}
\hline
w.o./with $E$ field & $\mathcal{P}$ & $\mathcal{T}$ & $\mathcal{PT}$   \\ \hline
$g_1 = -g_2 \neq 0$  & \ding{55}/\ding{55} & \ding{55}/\ding{55} & \ding{51}/\ding{55}  \\ \hline
$|g_{1}| \neq |g_{2}|$ & \ding{55}/\ding{55} & \ding{55}/\ding{55} & \ding{55}/\ding{55} \\ \hline
\end{tabular}
\end{table}

\textit{Layer Hall Effect in TI-Based Heterostructures---.}
In the absence of electric field, as demonstrated in Figs.~\ref{Fig1}(c) and \ref{Fig1}(f), the antiparallel arrangement of spins in magnetic layers, i.e., $g_{1}=-g_{2}$, induces doubly degenerate surface gaps, leading to an axion insulator with zero net Berry curvature due to the preservation of $\mathcal{PT}$ symmetry.
The further introduction of electric field breaks $\mathcal{PT}$ symmetry and gives rise to potential difference between the top and bottom surfaces, leading to an upward (downward) shift of the bands of the top (bottom) surface as displayed in Fig.~\ref{Fig1}(d). As a consequence, the Berry curvature carried by the surface bands has an energy shift (see Fig.~\ref{Fig1}(g)), resulting in an electric-field-induced LHE~\cite{Yi_2024}.

\begin{figure}[t]
  \centering
  \includegraphics[width=8.5 cm]{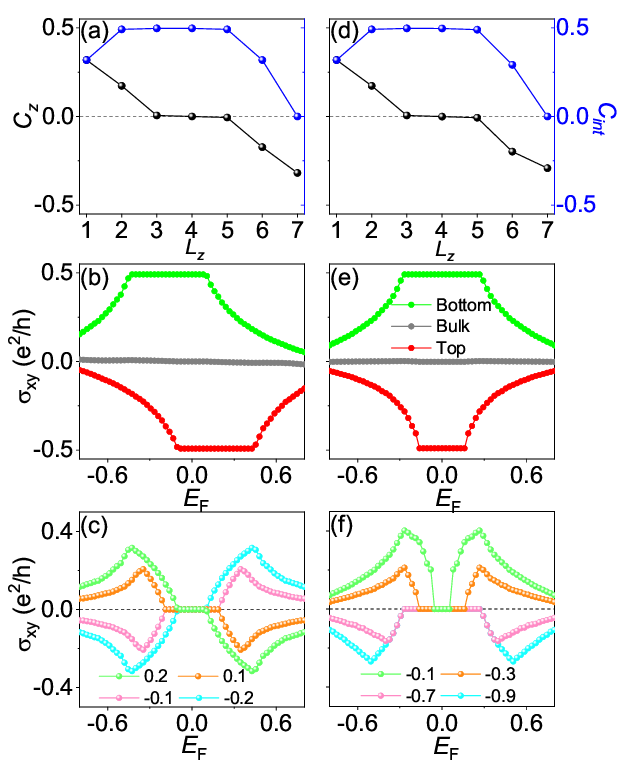}
  \caption{(a) Layer-resolved Chern number $\mathcal{C}_z$ and the cumulative Chern number $\mathcal{C}_{int}=\sum_{1}^{L_z}\mathcal{C}_z$ at $E_F = 0$, (b) surface-contributed and (c) total anomalous Hall conductivity $\sigma_{xy}$ as functions of Fermi level for the system with opposite exchange fields ($g_1 = - g_2$) in the presence of electric field. (d)-(f) Corresponding results for a system with inequivalent exchange fields in the absence of electric field. Different colors in (c) and (f) represent the strengths of electric field and exchange field $g_2$. Other parameters are consistent with those in Fig.~\ref{Fig1}.
}\label{Fig2}
\end{figure}

Previous studies have primarily focused on the special case with $g_{1}=-g_{2}$. However, when $|g_{1}| \neq |g_{2}|$, as displayed in Fig.~\ref{Fig1}(b), both Chern insulator and axion insulator phases can still be realized, e.g., the axion insulator with $\mathcal{C}=0$ always exists when $g_1 g_2 <0$. Realizing LHE requires the breaking of $\mathcal{PT}$ symmetry. As summarized in Tab.~\ref{table1}, the conventional approach to realize LHE with $g_{1}=-g_{2}\neq 0$ preserves (breaks) the $\mathcal{PT}$ symmetry in the absence (presence) of electric field. However, if $|g_{1}| \neq |g_{2}|$, the system simultaneously breaks $\mathcal{P}$, $\mathcal{T}$ and $\mathcal{PT}$ symmetries in the absence of electric field, which provides a more general approach to realize LHE by selecting inequivalent exchange field strengths between top and bottom magnetic layers. As displayed in Fig.~\ref{Fig1}(e), the inequivalent exchange fields open surface gaps with different amplitudes, leading to a splitting of surface bands. One can observe that both valence band maximum and conduction band minimum are from the top surface since $|g_{2}| < |g_{1}|$. This is different from the case with $g_{1}=-g_{2}$, as displayed in Fig.~\ref{Fig1}(d), where the two bands near Fermi level are from different surfaces. Consequently, the Berry curvature carried by the surface bands, as shown in Fig.~\ref{Fig1}(h), also differs from the former case presented in Fig.~\ref{Fig1}(g).

\begin{figure*}
  \centering
  \includegraphics[width=17.0 cm]{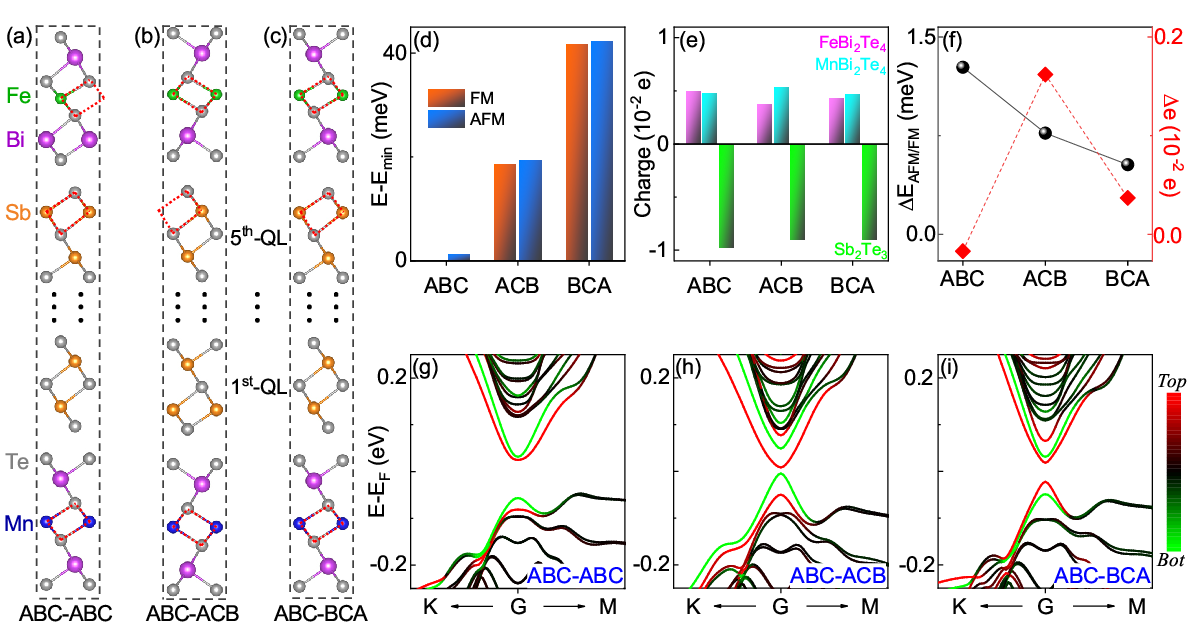}
  \caption{(a)-(c) Side views of the crystal structures of 1-SL MnBi$_2$Te$_4$/5-QL Sb$_2$Te$_3$/1-SL FeBi$_2$Te$_4$ with (a) ABC-ABC, (b) ABC-ACB, and (c) ABC-BCA stacking orders. (d) Energy difference $E-E_{min}$, (e) charge transfer, (f) magnetic coupling and charge transfer difference for the three configurations. (g)-(i) Layer-resolved band structures of the three heterostructures with (g) ABC-ABC, (h) ABC-ACB, and (i) ABC-BCA stacking orders, respectively, with green (red) color representing electronic states from the bottom (top) surface. Antiferromagnetic states are considered for the charge transfer and band structures.}\label{Fig3}
\end{figure*}

The distinct Berry curvature distributions in the two cases directly lead to different layer-resolved topological properties, as displayed in Fig.~\ref{Fig2}. For the electric-field-induced LHE, the nonzero $\mathcal{C}_z$ primarily originates from the outermost two layers of the systems, as shown in Fig.~\ref{Fig2}(a). In this case, the bottom and top two layers contribute approximately +0.5 and -0.5 to the Chern number, respectively. Therefore, we can define the anomalous Hall conductivity $\sigma_{xy}$ from the bottom/top two layers as bottom/top contribution, whereas that from the middle three layers as bulk contribution. As displayed in Fig.~\ref{Fig2}(b), the presence of electric field shifts the half-quantized $\sigma_{xy}$ from the bottom/top contributions downward/upward in energy space, whereas the contribution from the bulk is negligible. The shift of the surface anomalous Hall conductivity leads to a nonzero antisymmetric $\sigma_{xy}$ with respect to the Fermi level, as shown in Fig.~\ref{Fig2}(c). Moreover, the increase of electric field strength not only reduces the region where $\sigma_{xy}$ is zero but also enhances the maximum value of $\sigma_{xy}$.

In the presence of inequivalent exchange fields, e.g., ($g_1$, $g_2$)= (0.5, -0.2), the system exhibits distinct layer-resolved topological properties. As displayed in Fig.~\ref{Fig2}(d), the weaker $g_2$ leads to a smaller $\mathcal{C}_z$ in the top magnetic layer ($L_z =7$) and a larger $\mathcal{C}_z$ in the topological layer ($L_z =6$), compared with the case in Fig.~\ref{Fig2}(a). However, the sum of $\mathcal{C}_z$ from the top two layers still contributes to a half-quantized value. Due to the narrower gap of the top surface induced by the weaker $g_2$ as shown in Fig.~\ref{Fig1}(e), the in-gap half-quantized $\sigma_{xy}$ plateau for the top surface is smaller than that for the bottom surface in terms of energy, as illustrated in Fig.~\ref{Fig2}(e). Consequently, the nonzero anomalous Hall conductivity as a function of the Fermi level is observed, as displayed in Fig.~\ref{Fig2}(f). This indicates the realization of LHE by introducing inequivalent exchange fields into TI thin film. It is noteworthy that the size of zero plateau of $\sigma_{xy}$ shown in Fig.~\ref{Fig2}(f) is determined by the smaller exchange field strength between $g_1$ and $g_2$. For example, when $g_2$ varies from -0.1 to $-|g_1|$, the zero plateau of $\sigma_{xy}$ becomes wider due to the gradually increasing top surface gap. When $|g_2|$ exceeds $g_1$, the size of zero plateau remains constant due to the larger gap in the top surface. Furthermore, the sign of nonzero $\sigma_{xy}$ is determined by the larger exchange field between $g_1$ and $g_2$, e.g., when $|g_2| < |g_1|$ ($|g_2| > |g_1|$), a positive (negative) $\sigma_{xy}$ can be observed in Fig.~\ref{Fig2}(f).

The different topological origins of the two types of LHE result in a distinct form of anomalous Hall conductivity. Unlike the antisymmetric $\sigma_{xy}$ induced by electric field as shown in Fig.~\ref{Fig2}(c), the inequivalent exchange fields induced $\sigma_{xy}$ is symmetric with respect to $E_{F}=0$ as displayed in Fig.~\ref{Fig2}(f). This explicit difference of $\sigma_{xy}$ as a function of Fermi level between the two cases also provides a crucial fingerprint to distinguish the type of LHE in experiments.

\textit{Tuning the Type of Layer Hall Effect via Stacking Order---.}
To validate the theoretical analysis presented above, we perform first-principles calculations on realistic materials~\cite{SM}. Taking into account the well-established experimental techniques for fabricating TI thin films based on van der Waals materials, we construct heterostructures composed of a 5-QL Sb$_2$Te$_3$ coupled with one septuple layer (1-SL) of the magnetic insulators MnBi$_2$Te$_4$ and FeBi$_2$Te$_4$ on the bottom and top surfaces, respectively. As displayed in Figs.~\ref{Fig3}(a)-\ref{Fig3}(c), three types of stacking orders are considered, i.e., ABC-ABC, ABC-ACB, and ABC-BCA stackings [also see Fig. S1]. In all three configurations, the 5-QL Sb$_2$Te$_3$ is stacked in a standard ABC sequence, whereas the stacking order between magnetic monolayer MnBi$_2$Te$_4$/FeBi$_2$Te$_4$ and 5-QL Sb$_2$Te$_3$ can be tuned due to the weak van der Waals interactions. The ABC-BCA stacking can be obtained by an in-plane sliding of FeBi$_2$Te$_4$ from the ABC-ABC stacking, whereas the ABC-ACB stacking is distinct from the other two cases, as the MnBi$_2$Te$_4$/FeBi$_2$Te$_4$ layer follows an ACB sequence. While the monolayer FeBi$_2$Te$_4$ exhibits an in-plane magnetic easy axis~\cite{Wang_2023_2}, out-of-plane magnetization in FeBi$_2$Te$_4$ can be achieved by applying an external magnetic field. Therefore, we focus on the out-of-plane magnetization in the heterostructures.

Figure~\ref{Fig3}(d) shows the energy difference of the three configurations with ferromagnetic (FM) and antiferromagnetic (AFM) couplings. One can find that ABC-ABC stacking with FM coupling is the most energetically favorable configuration, whereas the total energies of the configurations with ABC-ACB and ABC-BCA stacking are higher by approximately 20 meV and 40 meV, respectively. For each configuration, the energy difference between AFM and FM states is less than 1.5 meV, as displayed in Fig.~\ref{Fig3}(f). Additionally, the magnetic materials MnBi$_2$Te$_4$ and FeBi$_2$Te$_4$ have different magnetic coercivities, suggesting that AFM states can be feasibly realized experimentally by applying a magnetic field. Hereinbelow, we focus on the AFM states of the heterostructures with different stacking orders.

The electronic wavefunction overlap between TI thin films and magnetic layers is influenced by the stacking order, leading to distinct layer-resolved electronic properties in the heterostructures with three configurations. For ABC-ABC configuration, as shown in Fig.~\ref{Fig3}(g), the bottom and top surfaces open gaps of approximately 88 meV and 106 meV, respectively. However, the valence band maximum is primarily contributed by bulk states along the high-symmetry line $\Gamma-M$, which hinders the realization of LHE. For the ABC-ACB stacking, as shown in Fig.~\ref{Fig3}(h), the bottom and top surfaces open gaps of approximately 53 meV and 59 meV, respectively, indicating that the strengths of the exchange fields induced in both surfaces of Sb$_2$Te$_3$ are similar. Furthermore, the low-energy bands near Fermi level are primarily contributed by the topological surface states, and their shape resembles that in Fig.~\ref{Fig1}(g), suggesting the presence of a strong built-in electric field. In the case of ABC-BCA stacking, as shown in Fig.~\ref{Fig3}(i), the bottom and top surfaces open gaps of approximately 80 meV and 41 meV, respectively, indicating that the exchange field strength induced in the bottom surface of Sb$_2$Te$_3$ is about twice as large as that in the top surface. Moreover, the shape of the low-energy bands is similar to that in Fig.~\ref{Fig1}(e), which suggests the presence of inequivalent exchange fields.

The origin of the strong built-in electric field in ABC-ACB configuration can be understood from differential charge density and Bader charge transfer analysis. First, the charge transfer process primarily occurs at interfaces of Sb$_2$Te$_3$/MnBi$_2$Te$_4$ and Sb$_2$Te$_3$/FeBi$_2$Te$_4$ (see Fig. S2~\cite{SM}). Second, as displayed in Fig.~\ref{Fig3}(e), electrons are transferred from Sb$_2$Te$_3$ to magnetic insulators MnBi$_2$Te$_4$ and FeBi$_2$Te$_4$ in all three configurations. In ABC-ABC and ABC-BCA configurations, the number of electrons transferred to MnBi$_2$Te$_4$ and FeBi$_2$Te$_4$ is similar. In contrast, for ABC-ACB configuration, more electrons are transferred from the bottom Sb$_2$Te$_3$ QL to MnBi$_2$Te$_4$. The charge transfer differences between MnBi$_2$Te$_4$ and FeBi$_2$Te$_4$ in the three configurations are explicitly demonstrated in Fig.~\ref{Fig3}(f). The smaller difference in ABC-ABC and ABC-BCA configurations indicates that the built-in electric field exerts a minor influence on the electronic properties. In contrast, the larger difference in ABC-ACB configuration leads to the formation of a strong built-in electric field.

\begin{figure}[t]
  \centering
  \includegraphics[width=8.5 cm]{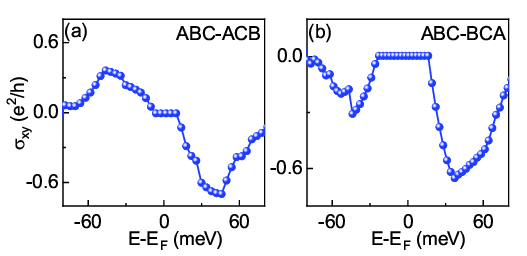}
  \caption{Anomalous Hall conductivity as a function of Fermi level for 1-SL MnBi$_2$Te$_4$/5-QL Sb$_2$Te$_3$/1-SL FeBi$_2$Te$_4$ heterostructures with (a) ABC-ACB and (b) ABC-BCA stacking orders.
}\label{Fig4}
\end{figure}

To further confirm LHE in the ABC-ACB and ABC-BCA configurations, we directly calculate the anomalous Hall conductivity $\sigma_{xy}$ as a function of Fermi level. As shown in Fig.~\ref{Fig4}, $\sigma_{xy}$ for ABC-ACB stacking is similar to that in Fig.~\ref{Fig2}(c), indicating the realization of LHE induced by a strong built-in electric field. In contrast, $\sigma_{xy}$ for the ABC-BCA stacking resembles that in Fig.~\ref{Fig2}(f), suggesting LHE is induced by inequivalent exchange fields. Therefore, different types of LHE, without requiring external electric field, can be realized in TI-based van der Waals heterostructures via tuning stacking orders.

\textit{Conclusion---.}
We demonstrate that LHE can be realized in heterostructures composed of magnetic insulators and TI thin films without requiring external electric field. In addition to the electric-field-induced LHE, we propose that inducing inequivalent exchange fields at the bottom and top surfaces of a TI thin film provides a distinct and crucial approach to realize LHE. By using van der Waals heterostructure composed of 1-SL MnBi$_2$Te$_4$/5-QL Sb$_2$Te$_3$/1-SL FeBi$_2$Te$_4$ as a realistic example, we show that both built-in electric-field-induced LHE and inequivalent exchange-field-induced LHE can be realized via tuning stacking orders between Sb$_2$Te$_3$ and MnBi$_2$Te$_4$/FeBi$_2$Te$_4$.

Our theoretical scheme is compatible with current experimental techniques for fabricating TI thin films based on van der Waals materials. Previous experiments have observed axion insulator states in sandwiched heterostructures composed of magnetic insulators separated by an undoped (Bi, Sb)$_2$Te$3$~\cite{Mogi_2017,Mogi_2017_2,Xiao_2018}, with nonzero $\sigma_{xy}$ measured when the electron density is far from the charge neutrality point~\cite{Mogi_2017}. This observation implies the presence of LHE without applying external electric field. To confirm the different types of LHE, further experimental exploration is needed. Our work provides a new approach to realize LHE without external electric field, and can stimulate further exploration of layer-resolved dissipationless electronics.

\begin{acknowledgments}
This work was financially supported by the National Natural Science Foundation of China (Grants No. 12474158 and 12488101), Anhui Initiative in Quantum Information Technologies (AHY170000), Innovation Program for Quantum Science and Technology (2021ZD0302800), the Natural Science Foundation of Fujian Province (No. 2022J05019), and China Postdoctoral Science Foundation (2023M733411 and 2023TQ0347). We are grateful to Supercomputing Center of University of Science and Technology of China for providing the high-performance computing resources.
\end{acknowledgments}


\end{document}